\documentclass[11pt]{article}
\usepackage{fa2025}
\usepackage{amsmath}
\usepackage{cite}
\usepackage{url}
\usepackage{graphicx}
\usepackage{color}
\usepackage{siunitx}
\usepackage[utf8]{inputenc}

\title{Acoustic Field Reconstruction in Tubes via Physics-Informed Neural Networks}

\multauthor
{Xinmeng Luan$^{1*}$ \hspace{1cm} Kazuya Yokota$^2$ \hspace{1cm} Gary Scavone$^1$} { 
  $^1$Computational Acoustic Modeling Laboratory, \\Center for Interdisciplinary Research in Music Media and Technology, \\ Schulich School of Music, McGill University, Canada\\
$^2$ Department of Mechanical Engineering, Nagaoka University of Technology, Japan\\
\correspondingauthor{xinmeng.luan@mail.mcgill.ca}{Luan et al.}
}

\begin{document}

\maketitle
\begin{abstract}
This study investigates the application of Physics-Informed Neural Networks (PINNs) to inverse problems in acoustic tube analysis, focusing on reconstructing acoustic fields from noisy and limited observation data. Specifically, we address scenarios where the radiation model is unknown, and pressure data is only available at the tube's radiation end. A PINNs framework is proposed to reconstruct the acoustic field, along with the PINN Fine-Tuning Method (PINN-FTM) and a traditional optimization method (TOM) for predicting radiation model coefficients. The results demonstrate that PINNs can effectively reconstruct the tube's acoustic field under noisy conditions, even with unknown radiation parameters. PINN-FTM outperforms TOM by delivering balanced and reliable predictions and exhibiting robust noise-tolerance capabilities.

\end{abstract}
%

\section{Introduction}\label{sec:introduction}

Physics-Informed Neural Networks (PINNs) have recently garnered significant attention in computational physics \cite{raissi2019physics}.
 By incorporating underlying physical laws, typically expressed as ordinary or partial differential equations (ODEs or PDEs), into the training process, PINNs can effectively model the behavior of physical systems, even in scenarios with limited or noisy data.
Given PINNs' exceptional performance in solving inverse problems and their robustness to noise, they offer a promising new approach to tackling physics-related engineering challenges.
PINNs have been applied to various acoustics inverse problems, such as near-field acoustic holography \cite{olivieri2021physics, luan2024complex}, acoustic boundary admittance estimation \cite{schmid2024physics} and sound field reproduction \cite{karakonstantis2024room}.

Time-domain acoustic analysis of wind instruments generally involves solving the 1D plane wave equation within tubes. Traditional numerical methods, such as the Finite Difference Method (FDM) and the Finite Element Method (FEM), are commonly used to solve the wave equation. Recently, Yokota et al. proposed a PINNs framework to address the time-domain 1D plane acoustic wave equation in tubes \cite{yokota2024physics, yokota2024synthesis, yokota2024cnn, yokota2024identification}. Their work includes analyzing both the forward problem with the application of sound synthesis, which involves solving the PDE \cite{yokota2024physics, yokota2024synthesis}, and the inverse problem, focused on identifying energy loss coefficients \cite{yokota2024physics, yokota2024identification} and tube geometry parameters \cite{yokota2024physics, yokota2024cnn} using observed pressure data at the tube's radiation end. 
However, the study revealed some limitations: 
While the inverse problem of identifying system parameters showed promising results for geometric parameters \cite{yokota2024physics, yokota2024cnn}, it did not perform as well for estimating energy loss coefficients \cite{yokota2024physics, yokota2024identification}. For the inverse problem \cite{yokota2024physics, yokota2024identification, yokota2024cnn, yokota2024synthesis}, they treated the desired parameters or coefficients as trainable neural network parameters and optimized them simultaneously with the neural network by minimizing the loss function.


In this paper, we aim to advance the application of PINNs to acoustic tube problems, with a particular emphasis on inverse problems. Specifically, we address the reconstruction of the acoustic field within tubes using PINNs, based on noisy and limited observation data. 
Our focus is on scenarios where the radiation model is unknown, and pressure data is available only at the tube's radiation end. The radiation model is considered unknown because, in practice, it depends on the geometry of the structure and is typically difficult to determine accurately. While time-domain radiation models exist for simpler cases, such as infinite flanged or unflanged terminations, determining the appropriate model becomes challenging for more complex geometries like bells \cite{silva2009approximation}. The problem is even more pronounced in cases like the two axially distributed front-end holes in the \textit{dizi} \cite{luan2022acoustical}, for which no time-domain radiation model has been derived.
Additionally, we propose PINN Fine-Tuning Method (PINN-FTM) and traditional Optimization Method (TOM) to tackle the prediction of the radiation model coefficients.



\section{Problem formulation}
\subsection{Governing equation and conditions}
The time domain 1D plane acoustic wave propagation in an axisymmetric 1D lossy tube of length $L$ with vary cross-section $A(x)$ can be characterised by the second-order horn equation \cite{yokota2024physics}
\begin{equation}
\begin{aligned}
	 & \phi_{xx} + \frac{1}{A} A_x \phi_{x} = GR \phi + \bigg (\frac{G\rho}{A} + \frac{RA}{K} \bigg) \phi_t + \frac{\rho}{K} \phi_{tt}, \\ & \qquad \qquad \qquad \qquad  \qquad \qquad x \in [0,L],  \quad  t \in [0,T]
    \end{aligned}
\end{equation}
where $\phi$ represents the velocity potential, while $G$ and $R$ denote the coefficients of energy loss due to thermal conduction and viscous friction at the tube wall, respectively. $K$ is the bulk modulus, and $\rho$ is the air density. $T$ represents  the wave period.
The relationships between the sound pressure $p$, the volume velocity $u$, and the velocity potential $\phi$ are given by
\begin{equation}
\begin{cases}
\displaystyle{ p = RA\phi + \rho \phi_t, }\\ \displaystyle{
    u = -A \phi_x.}
\end{cases}	
\end{equation}
The following thermal and viscous loss model are employed, under the assumptions that the wall surface is rigid and thermal conductivity is infinite, expressed by \cite{yokota2024physics}
\begin{equation}
    \begin{aligned}
        R &= \frac{S}{A^2} \sqrt{\frac{\omega_c \rho \mu}{2}}, \\
        G &= S \frac{\eta - 1}{\rho c^2} \sqrt{\frac{\lambda \omega_c}{2 c_p \rho}},
    \end{aligned}
\end{equation}
with $S$ the circumference of the acoustic tube, $\mu$ the viscosity coefficient, $\eta$ the heat-capacity ratio, $c$ the speed of sound, $\lambda$ the thermal conductivity and $c_p$ the specific heat at constant pressure, and $\omega_c$ the angular velocity for the energy loss term. Alternatively, a more advanced model incorporating higher-order thermal and viscous losses proposed in \cite{thibault2021dissipative} can be utilized for improved accuracy.

The radiation model at the tube open end can be expressed as \cite{thibault2020time}
\begin{equation}
    \frac{\rho c}{A} u_t= \alpha p + \beta p_t, \quad x=L, \quad  t \in [0,T],
    \label{eq: radiation}
\end{equation}
where the coefficients $\alpha$ and $\beta$ are related to the frequency-domain radiation impedance, expressed using a second-order Taylor series expansion \cite{chaigne2016elementary}
\begin{equation}
    Z_r = \frac{\rho c}{\pi r^2} \left(\delta j kr + \beta_c (k r)^2\
 \right),
 \label{eq: taylor}
\end{equation}
with $r$ the opening radius, $k=\omega/c$ the wavenumber, and $\omega$ the angular frequency.
Equation~\eqref{eq: taylor} can be reformulated into a first-order Padé development as \cite{rabiner1978digital}
\begin{equation}
     Z_r = \frac{\rho c}{\pi r^2} \frac{jkr}{\alpha + jkr \beta},
\end{equation}
with 
\begin{equation}
        \alpha = \frac{1}{\delta}, \quad 
        \beta = \frac{\beta_c}{\delta^2}.
\end{equation}

At the other end, the input wave is imposed as a boundary condition 
\begin{equation}
    u = u \big|_{x=0}, \quad t \in [0,T].
\end{equation}
A periodicity condition is applied, as
\begin{equation}
    \phi \big |_{t=0} = \phi \big |_{t=T}, \quad x \in [0,L].
\end{equation}

\subsection{Inverse problem statement}
\begin{figure}
    \centering
    \includegraphics[width=1\linewidth]{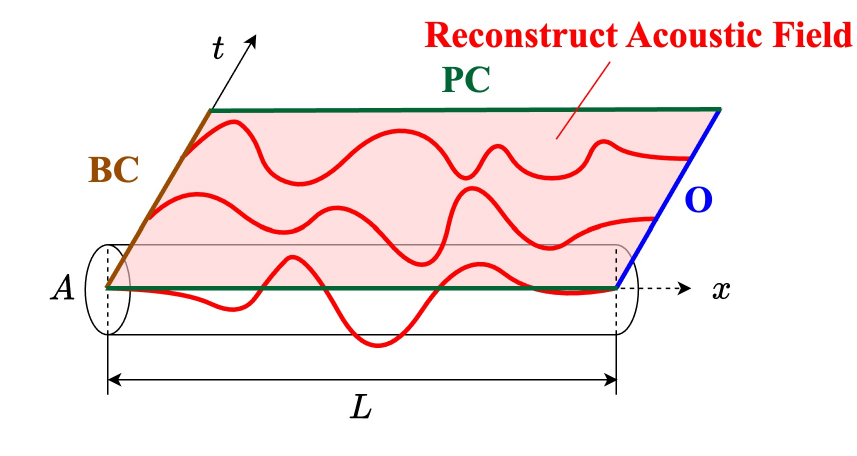}
    \caption{Schematic diagram of acoustic field reconstruction in a tube.}
    \label{fig: tube}
\end{figure}

Assuming the expression for the time-domain radiation model (\ref{eq: radiation}) is unknown, the primary objective of the inverse problem is to reconstruct the time-space acoustic field along the tube, as shown in Fig.~\ref{fig: tube}.
Subsequently, the second goal is to predict the radiation model using pressure observation data at the boundary $x=L$.

\section{Proposed Method}

\subsection{Acoustic field reconstruction via PINNs}\label{sec: 3.1}
\begin{figure*}
    \centering
\includegraphics[width=0.8\linewidth]{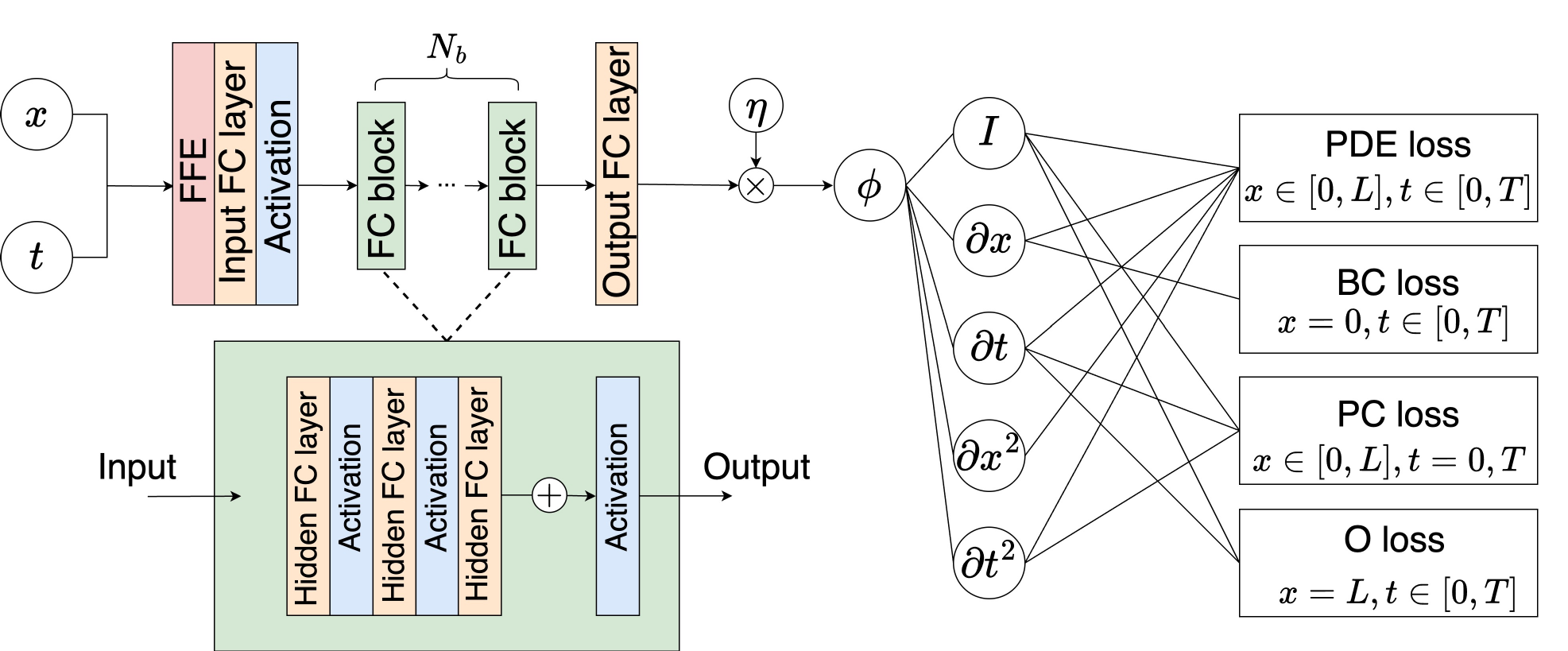}
    \caption{PINN architecture.}
    \label{fig:nn_arch}
\end{figure*}

A deep residual neural network is utilized, taking $x$ and $t$ as inputs and producing $\hat{\phi}(x,t)$ as the output
\begin{equation}
   \hat{\phi} (x,t) = \Gamma (\gamma, x, t),
   \label{eq: phi_formu}
\end{equation}
where $\Gamma$ represents an estimator modeled by the residual neural network, and $\gamma$ are the trainable neural network parameters.
The neural network architecture is similar to that in \cite{yokota2024physics}, as shown in Fig.~\ref{fig:nn_arch}. However, thanks to the radiation formulation in Eq.~\eqref{eq: radiation}, only a single trunk is used, rather than an additional trunk for separately predicting the radiation, as in \cite{yokota2024physics}.
The activation function used is Snake \cite{ziyin2020neural}, which has been reported to be robust for periodic inputs, consistent with \cite{yokota2024physics}. In Fig.~\ref{fig:nn_arch}, the Input Fully Connected (FC) layer consists of 200 input channels and $N_f$ output channels. The hidden FC layers have $N_f$ channels for both input and output, while the Output FC layer comprises $N_f$ input channels and a single output channel. There are $N_b$ FC blocks in total. The inputs $x$ and $t$ are normalized to the range $[-1, 1]$ to facilitate easier training, while the output $\phi$ is scaled by $\xi$ to return to its actual range. Moreover, the random  Fourier Feature Embedding (FFE) \cite{tancik2020fourier} is employed with a scale parameter $\sigma$ set to 0.1, which controls the range of frequencies in the embedding, and is applied with an encoding size of 50. FFE is a technique that maps input data into a higher-dimensional space using sinusoidal transformations, which has been found to effectively enhance the training accuracy of PINNs. The value $\sigma = 0.1$ is selected based on experimental results. While this is a relatively low value, higher values can lead to the failure of neural network training. This indicates that the utilization of Fourier feature embedding in this case may not be strictly necessary.

Automatic differentiation is then used to compute the loss functions, which encompass the PDE, boundary condition (BC), periodicity condition (PC), and observation (O) losses, all formulated as mean squared error (MSE) terms. 
The PDE loss is expressed as
\begin{equation}
\begin{aligned}
    \mathcal{L}_{PDE} &= \frac{1}{N_{PDE}} \bigg \| \hat{\phi}_{xx}
     + \frac{1}{A} A_x \hat{\phi}_{x} - GR \hat{\phi} - \\ & \qquad \qquad  \qquad   \bigg (\frac{G\rho}{A} -   \frac{RA}{K} \bigg) \hat{\phi}_t - \frac{\rho}{K}  \hat{\phi}_{tt}  
    \bigg \|_2^2, \\  & \qquad \qquad \qquad  \qquad \qquad x \in [0,L],  \quad  t \in [0,T].
    \end{aligned}
\end{equation}
The BC loss is 
\begin{equation}
    \mathcal{L}_{BC} = \frac{1}{N_{BC}} \big \| \hat{u} - u \big \|_2^2, \quad x =0, \quad t \in [0,T].
\end{equation}
The PC loss accounts for the periodicity of  $u$, $p$, and the second time derivative of  $\phi$, as
\begin{equation}
\setstretch{1.5}
\begin{cases} \displaystyle{
     \mathcal{L}_{PC, u} = \frac{1}{N_{PC}} \big \| \hat{u}\big|_{t=0} - \hat{u}\big |_{t=T} \big \|_2^2,} \\  \displaystyle{
     L_{PC, p} = \frac{1}{N_{PC}} \big  \| \hat{p}\big|_{t=0} -  \hat{p}\big|_{t=T} \big \|_2^2,} \\ \displaystyle{
    L_{PC, \phi_{tt}} = \frac{1}{N_{PC}} \big \| \hat{\phi}_{tt}\big|_{t=0} - \hat{\phi}_{tt}\big|_{t=T} \big \|_2^2.}
\end{cases}
x \in [0,L].
\end{equation}
The O loss is
\begin{equation}
    \mathcal{L}_{O} = \frac{1}{N_{O}} \big \| \hat{p} - p \big \|_2^2, \quad x =L, \quad t \in [0,T].
\end{equation}
$N_{PDE}$, $N_{BC}$, $N_{PC}$ and  $N_{O}$ represent the respective numbers of collocation points used for the PDE, BC, PC, and O loss computations.
Then the total loss function is 
\begin{equation}
\begin{aligned}
    \mathcal{L} &= \lambda_{PDE} \mathcal{L}_{PDE} + \lambda_{BC} \mathcal{L}_{BC} + \lambda_{O} \mathcal{L}_{O} +\\ & \lambda_{PC} (\lambda_{PC,u} \mathcal{L}_{PC,u}  + \lambda_{PC,p} \mathcal{L}_{PC,p}  + \lambda_{PC,\phi_{tt}} \mathcal{L}_{PC,\phi_{tt}}),
\end{aligned}
\end{equation}
with $\lambda_{PDE} $, $ \lambda_{BC}$, $ \lambda_{O}$, $ \lambda_{PC}$, $\lambda_{PC,u}$ and $ \lambda_{PC,p}$ as the loss function weights.

\subsection{Radiation model reconstruction}
To reconstruct the radiation model, we propose two approaches. Assuming the radiation model formulation \eqref{eq: radiation} is known, the first approach, the PINN Fine-Tuning Method (PINN-FTM) involves fine-tuning the neural network parameters $\gamma$. In this method, the neural network is initialized with the parameters of the trained PINN. Additionally, two trainable parameters, the radiation coefficients $\alpha$ and $\beta$, are optimized simultaneously during the PINN training. FTM can also be viewed as a transfer learning procedure.
Therefore, the formulation in \eqref{eq: phi_formu} becomes
\begin{equation}
   \hat{\phi} (x,t) = \Lambda (\gamma, \alpha, \beta, x, t),
\end{equation}
where $\Lambda$ represents an estimator modeled by this neural network.
For simplicity, we refer to the PINN described in Sec.~\ref{sec: 3.1} as $\Gamma$ and the fine-tuned PINN as $\Lambda$, based on their respective parameter notations.

In the second approach, the Traditional Optimization Method (TOM), we use automatic differentiation to obtain $\displaystyle{\hat{u}_t}$, $\displaystyle{\hat{p}}$, and $\displaystyle{\hat{p}_t}$ at $x=L$ after the PINN $\Gamma$ is trained. With the known formulation of the radiation model \eqref{eq: radiation}, the coefficients $\alpha$ and $\beta$ are then determined through an optimization procedure.
\begin{equation}
\begin{aligned}
     \hat{\alpha}, \hat{\beta} = \arg\min_{\alpha, \beta} \left\| \frac{\rho c}{A} \hat{u}_t - \alpha \hat{p} - \beta \hat{p}_t \right\|_2^2, \\
    x = L, \quad t\in [0,T].
\end{aligned}
\label{eq: opt}
\end{equation}

\section{Validation of the proposed method}

\subsection{Implementation}

The neural network has $N_f = 200$ and  $N_b = 5$ , with a total of 643,401 trainable parameters.
The numbers of collection points are 
$N_{PDE} = 5000$ and $N_{BC} =N_{PC} =N_{O} = 1000$. The data points for the PDE are randomly generated using a 2D Sobol sequence, while the remaining points are uniformly sampled along their domain.
The manually chosen loss function weights are listed in Table~\ref{tab:lambda}.

\begin{table}[h!]
    \centering
    \begin{tabular}{c c c c }
     \hline
   $\lambda_{PDE}$  & $\lambda_{BC}$ & $\lambda_{O}$ & $\lambda_{PC}$ \\
     $5 \times 10 ^{-6}$ & $3.4 \times 10 ^{5}$& 1 & 1 \\
      \hline
     $\lambda_{PC, u}$ & $\lambda_{PC, p}$ & $\lambda_{PC, \phi_{tt}}$ & \\
      $5 \times 10 ^{4}$ & 1 & $1 \times 10 ^{-8}$& 
      \\ \hline
\end{tabular}
    \caption{Loss function weights.}
    \label{tab:lambda}
\end{table}

We choose the tube length  $L= 1$, and the frequency $f = \SI{261.6}{\hertz}  $ with the period $T=1/f$, as the test case, which aligns with \cite{yokota2024physics}.
The infinite flanged case is employed for the radiation model, with $\delta = 0.8236$, $\beta_c = 0.5$.
A smoothed Rosenberg volume velocity waveform is used at $x=0$ for the test case, which simulates a glottal pulse, in alignment with \cite{yokota2024physics}. 
The observed pressure data at $x=L$ is obtained using the Finite Difference Method (FDM), centered-time centered-space scheme, similar to \cite{yokota2024physics}, with a special treatment for the radiation condition employing a predictor-corrector iterative scheme. 
To simulate a more challenging scenario, \SI{40}{dB} signal-to-noise ratio (SNR) Gaussian noise is added to the signal, mimicking observations obtained from a microphone.
\begin{figure*}[h!]
    \centering
    \includegraphics[width=0.9\linewidth]{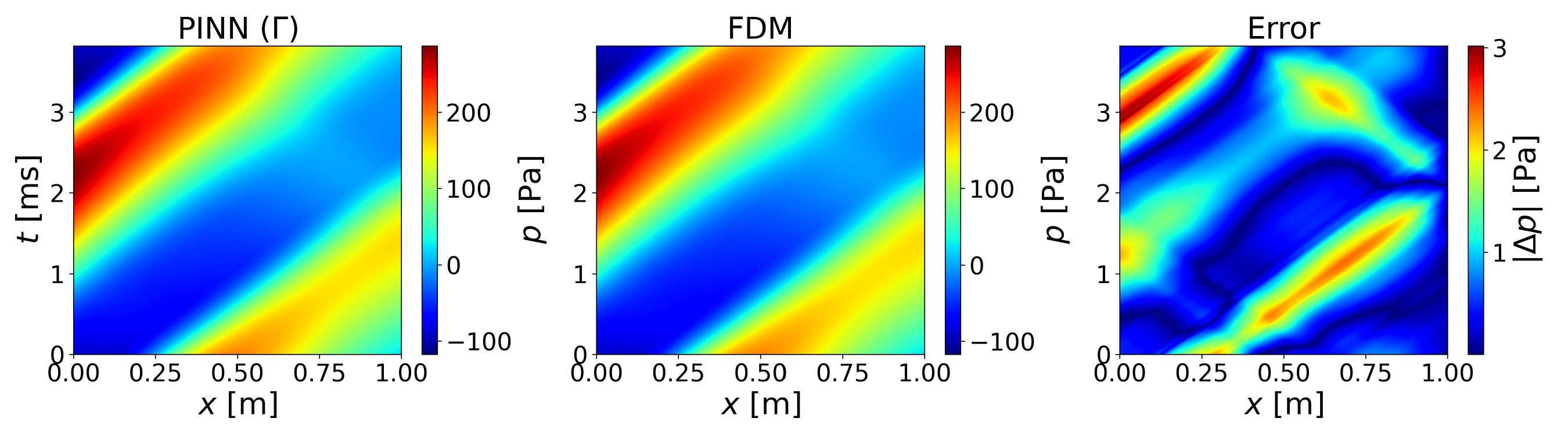}
    \caption{Time-space pressure field along the tube: results from PINN ($\Gamma$) (left), FDM (middle), and the error (absolute difference) between PINN and FDM (right).}
    \label{fig: p_all}
\end{figure*}

\begin{figure*}
    \centering
    \includegraphics[width=1\linewidth]{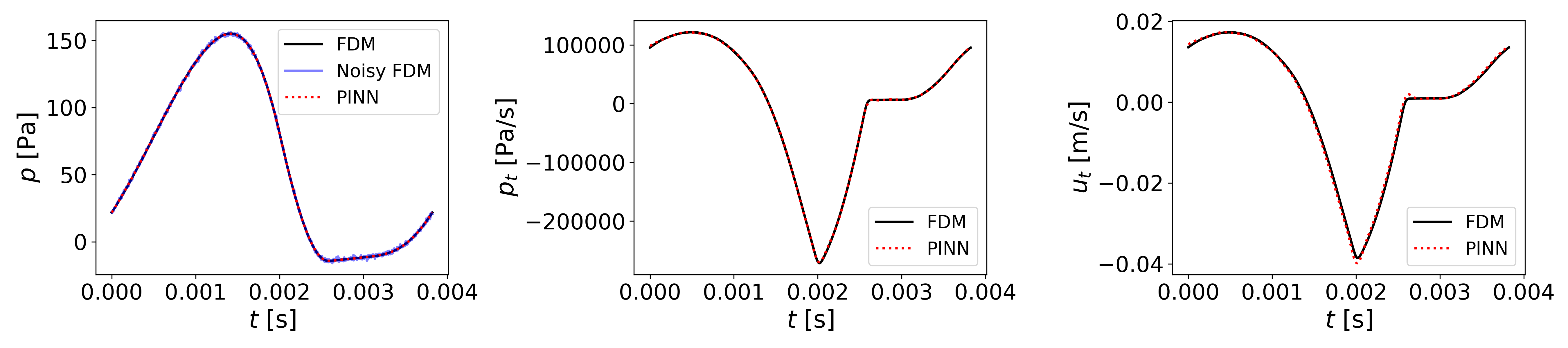}
    \caption{$p$, $p_t$ and $u_t$ at $x=L$, from PINN ($\Gamma$).}
    \label{fig: p_end}
\end{figure*}

We proceed by training the resulting PINN model via full-batch gradient descent.
The training procedure for $\Gamma$ utilizes the Adam optimizer \cite{kingma2014adam} to optimize $\gamma$, incorporating a learning rate decay scheme as described in \cite{yokota2024physics} 
\begin{equation}
    \lambda_{adam}(i_e) = \frac{ \lambda_{adam, init}}{1 + 0.007 i_e},
    \label{eq: adam}
\end{equation}
where $i_e = 20000$ represents the total number of epochs, and $\lambda_{adam, init} = 1 \times 10^{-2}$ is the initial learning rate. To further improve accuracy, a second-order L-BFGS optimizer \cite{liu1989limited} is employed for an additional 3000 epochs following the Adam optimization. The L-BFGS optimizer is configured with a learning rate of 1, a maximum of 20 iterations and a history size of 10.  
For $\Lambda$, two optimizers are employed. The primary optimizer, an Adam optimizer, is used to optimize $\gamma$, following the same learning rate decay scheme in \eqref{eq: adam}, with $\lambda_{adam, init} = 1 \times 10^{-4}$ and $i_e = 20000$. A secondary Adam optimizer is used to optimize the radiation coefficients $\alpha$ and $\beta$, also following the decay scheme in \eqref{eq: adam}, with $\lambda_{adam, init} = 1 \times 10^{-2}$ and $i_e = 20000$. The initial values for $\alpha$ and $\beta$ are both set to 1. 
The use of two different optimizers, with the primary optimizer having a smaller learning rate than the secondary optimizer, reflects the principle of fine-tuning. This approach ensures that the parameters $\gamma$ undergo minimal changes during training, preserving its pre-trained structure while allowing $\alpha$ and $\beta$ to adapt more significantly.
After this initial optimization, a second-order L-BFGS optimizer is applied for an additional 1500 epochs to refine the parameters further.

\textit{Pytorch} is used for PINNs implementation.
FFE is implemented using package \cite{long2021rffpytorch}.
In the TOM, the quasi-Newton method is implemented using MATLAB's \texttt{optimoptions} function to carry out the optimization problem described in \eqref{eq: opt}.
In the TOM, the problem described in \eqref{eq: opt} is solved using MATLAB's \texttt{lsqnonlin} function, which performs nonlinear least squares optimization based on the trust-region-reflective algorithm.
The lower and upper bounds for $\alpha$ and $\beta$  are both set within the range $[0,5]$.
 The code is available as open source \footnote{https://github.com/Xinmeng-Luan/PINNtube\_inverse\_fa25}. 

\subsection{Results}

\subsubsection{Acoustic field reconstruction: $\Gamma$}
To evaluate the PINN's performance in reconstructing the space-time sound pressure within the tube, we tested the model using $x$ and $t$ as inputs sampled on an evenly spaced grid. The grid consists of $5001$ points along $x$ and $1001$ points along $t$, matching the resolution used in the FDM for direct comparison.
The reconstructed acoustic field in the tube is shown in Fig.~\ref{fig: p_all}. The time-space acoustic field features are clearly visible, demonstrating that the PINNs are capable of accurately reconstructing the acoustic field in the tube, even without knowledge of the radiation model, using noisy and limited observation data.

To better understand the predicted acoustic data at the tube's radiation end, Fig.~\ref{fig: p_end} compares the visualization of $p$, $p_t$, and $u_t$ at $x=L$ from PINN ($\Gamma$) with the FDM results. Overall, the shapes of the three curves align well with the FDM data, with $p$ showing strong agreement, while $p_t$ and $u_t$ exhibit weaker alignment. This discrepancy contributes to TOM's failure to accurately predict $\alpha$. 
Notably, the error in $\alpha$ is consistently larger than that in $\beta$ across all approaches (PINN-FTM, TOM, and TOMB) in Table\ref{tab: rad}. This trend may be attributed to the optimization formulation in \eqref{eq: opt}, where $p_t$ exhibits a significantly larger order of magnitude than $p$, as shown in Fig.~\ref{fig: p_end}. Consequently, since $\alpha p$ is relatively small compared to $\beta p_t$, a larger error in $\alpha$ may be more permissible.

\subsubsection{Radiation model reconstruction: $\Lambda$}

The predicted radiation model coefficients from PINN-FTM and TOM are presented in Table~\ref{tab: rad}. Additionally, we evaluate the efficiency of the TOM method using accurate, noise-free $p$, $p_t$, and $u_t$ data from FDM, with the corresponding predicted results  (referred to as TOM Baseline, TOMB)  also included in Table~\ref{tab: rad}. 
The results show that PINN-FTM offers balanced and reliable predictions for both $\alpha$ and $\beta$. In contrast, TOM achieves greater accuracy for $\beta$ but suffers from significant instability in $\alpha$ estimation. 
However, with noise-free data, TOMB performs well, indicating its strong dependence on ideal data conditions for reliable results. 
Overall, this suggests that PINN-FTM is more suitable for inverse radiation model prediction when working with noisy observational data.

\begin{table}[]
    \centering
    \begin{tabular}{c c c c c}
    \hline
     & PINN-FTM & TOM & TOMB  & GT\\
     \hline
       $\alpha$  & 1.1039 &4.4409e-14& 1.2161 & 1.2142\\
       $\beta$ & 0.7811 & 0.7435  & 0.7371 & 0.7371 \\
       \hline
    \end{tabular}
    \caption{Comparison of $\alpha$ and $\beta$ obtained from PINN-FTM, TOM, TOMB, and Ground Truth (GT).}
    \label{tab: rad}
\end{table}

\begin{figure}
    \centering
    \includegraphics[width=1\linewidth]{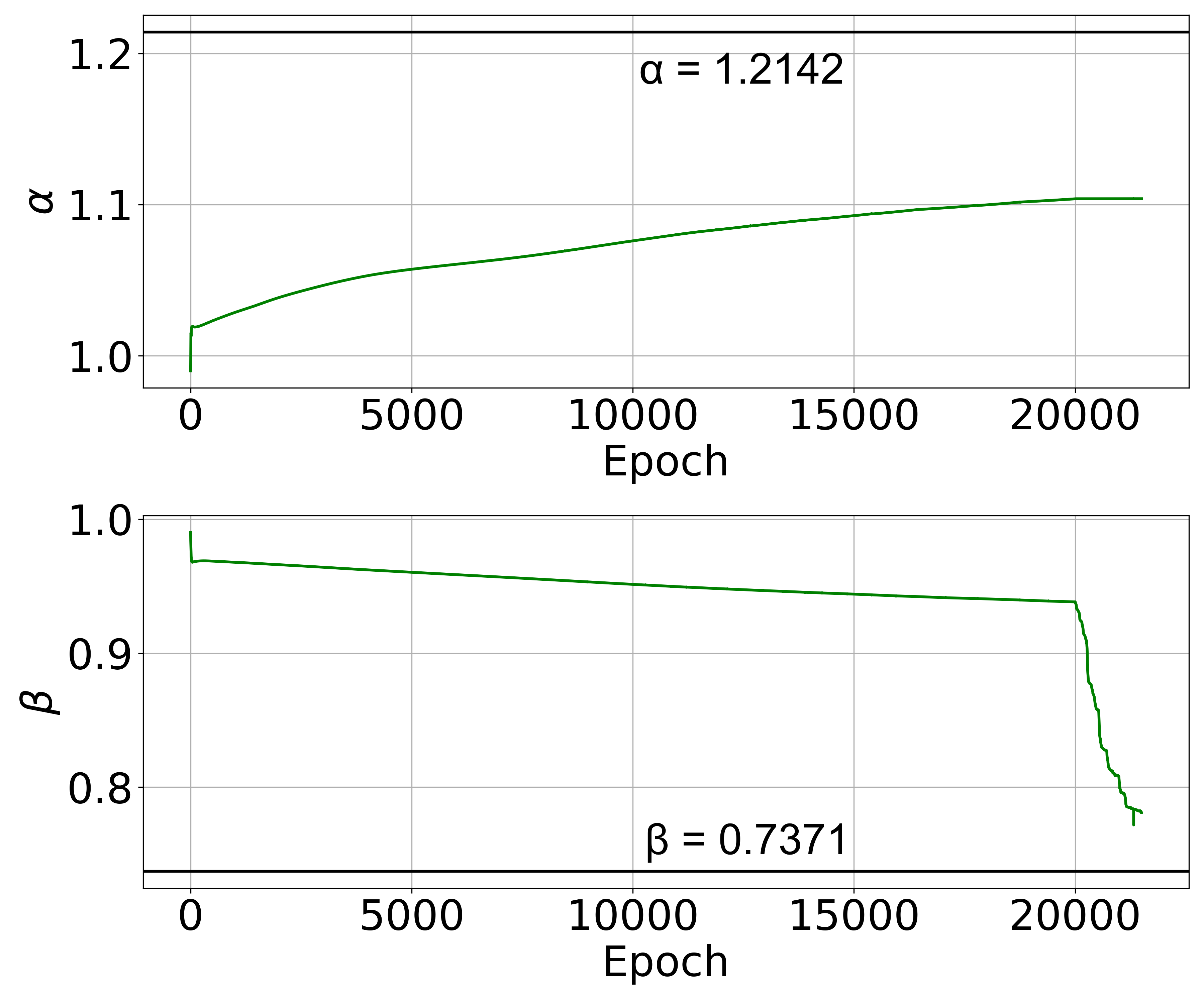}
    \caption{The PINN-FTM ($\Lambda$) training progress of $\alpha$ and $\beta$ over epochs.}
    \label{fig: alpha_beta}
\end{figure}

To further evaluate PINN-FTM, the training progress of $\alpha$ and $\beta$ over epochs is visualized in Fig.~\ref{fig: alpha_beta}. The results indicate that during Adam optimization, both $\alpha$ and $\beta$ gradually converge toward the ground truth. The second-order L-BFGS optimizer further accelerates the optimization of $\beta$ but has minimal effect on $\alpha$, highlighting the advantage of second-order optimization methods in improving convergence.

\subsection{Discussion}

We have demonstrated the effectiveness of the proposed PINN in reconstructing the tube’s acoustic field using noisy pressure observation data at the radiation end, as well as the capability of PINN-FTM in predicting the radiation model. However, there are certain limitations that we have not addressed in this paper. Since our primary aim is to highlight the feasibility of this framework, we have only tested the method under periodic boundary conditions, which inherently correspond to predicting the steady state. We would like to emphasize that PINNs, in general, are not limited to this type of problem. They can be applied to scenarios with initial condition constraints and can also handle long-time problems, although this may require the use of a time-marching scheme \cite{krishnapriyan2021characterizing} or other strategies.

Furthermore, this study has solely relied on simulation data. Testing the approach with real-world experimental data would be valuable, although we expect potential challenges, such as microphone placement. Certain settings within the proposed approach may need careful adjustments to better suit real-world conditions. For instance, PINN could be tailored for impedance tube measurements, which are much more common in acoustic measurements and involve a larger number of microphones for observation data collection.

The proposed technology can be applied to duct acoustics and acoustic waveguide problems.
While our primary interest lies in the potential application of inferring the acoustic field in wind musical instruments, acoustic tubes are also widely used in various engineering applications, such as impedance tube measurements, biomedical acoustics (including vocal tract modeling), and aeroacoustics. For instance, acoustic tubes play a key role in thermoacoustic engines, where reconstructing the acoustic field within the tube is essential \cite{kuzuu2015reconstruction}.
A significant advantage of the PINN is its ability to operate in the time domain, while most traditional acoustic tube measurements are limited to the frequency domain analysis \cite{kuzuu2015reconstruction}.
Moreover, our work can also be connected to the sound field reconstruction problem in room acoustics using PINNs, as explored in \cite{karakonstantis2024room}, where the objective is to predict the sound field in the room from sparse pressure measurement data, without predefined boundary conditions, and governed by the inhomogeneous wave equation. 
More generally, this acoustic field reconstruction goal can be framed as an inverse problem for PINNs, where the boundary conditions are not known a priori, and the objective is to recover unknown parameters or functions from limited and noisy observations. For a comprehensive discussion on the application of PINNs to such problems, see \cite{mishra2022estimates}.

\section{Conclusion}
In this study, we propose a PINNs framework for reconstructing the acoustic field in tubes and introduce the PINN Fine-Tuning Method (PINN-FTM) alongside a traditional optimization method (TOM) to predict the radiation model coefficients. The results demonstrate that PINNs effectively reconstructs the tube's acoustic field using noisy pressure observation data at the radiation end, even with unknown radiation conditions. Compared to TOM, PINN-FTM delivers balanced and reliable predictions, showcasing its noise-tolerance capability, providing a novel approach for retrieving time-domain radiation models using limited measured data.
This research explores a potential application of inverse problems in acoustic tube analysis, showing promise for future testing and development with real-world experimental data.





\bibliography{fa2025_template}

%
%
%

\end{document}